# Riroriro: Simulating gravitational waves and evaluating their detectability in Python

**Wouter G. J. van Zeist[1], Héloïse F. Stevance[1], and J. J. Eldridge[1]**

**1** Department of Physics, University of Auckland, New Zealand

## Summary

`Riroriro` is a Python package to simulate the gravitational waveforms of binary mergers of black holes and/or neutron stars, and calculate several properties of these mergers and waveforms, specifically relating to their observability by gravitational wave detectors.

The gravitational waveform simulation of `Riroriro` is based upon the methods of Buskirk & Babiuc-Hamilton (2019), a paper which describes a computational implementation of an earlier theoretical gravitational waveform model by Huerta et al. (2017), using post-Newtonian expansions and an approximation called the implicit rotating source to simplify the Einstein field equations and simulate gravitational waves. `Riroriro`'s calculation of signal-to-noise ratios (SNR) of gravitational wave events is based on the methods of Barrett et al. (2018), with the simpler gravitational wave model `Findchirp` (Allen et al., 2012) being used for comparison and calibration in these calculations.

## Statement of Need

The field of gravitational wave astronomy has been especially active since the first observation of gravitational waves was announced in 2016 (Abbott et al., 2016). Observations of gravitational waves from binary mergers can provide unique information about their progenitors and stellar populations, especially when combined with electromagnetic observations in the field called multi-messenger astronomy. A major factor in the successful detection and analysis of gravitational wave signals is the creation of simulations of such signals which observed data can be compared to. Because of this, multiple gravitational wave models have been created over the years. In particular, the gravitational wave observatories LIGO and Virgo have created their own models to use as templates in gravitational wave searches, with the main software for this being `LALSuite` (LIGO Scientific Collaboration, 2018). Various research groups have created waveform models, with some examples of recent sophisticated waveform models being `IMRPhenomXPHM` (Pratten et al., 2020) and `SEOBNRv4PHM` (Ossokine et al., 2020), which are both also available in `LALSuite`.

We have not tested if the waveform model of `Riroriro` that is based on Huerta et al. (2017) and Buskirk & Babiuc-Hamilton (2019) is accurate enough to use for parameter estimation of detected gravitational wave transients as this was not within the scope of our project; it is likely that accurate parameter estimation requires careful modelling of the ringdown phase, especially for the most massive mergers. However, we use a level of accuracy adequate for our aim of modelling the detectability of gravitational wave transients predicted by stellar population syntheses. Furthermore, the code of `Riroriro` is structured and commented in such a way that each step in the process of the simulation is individually identifiable and modifiable by users. Users could also substitute in functions from other algorithms or even use the detectability modules on waveforms from other sources, as long as the user puts these in Riroriro's format.

`Riroriro` combines areas covered by previous models (such as gravitational wave simulation, SNR calculation, horizon distance calculation) into a single package with broader scope and versatility in Python, a programming language that is ubiquitous in astronomy. Aside from being a research tool, `Riroriro` is also designed to be easy to use and modify, and it can also be used as an educational tool for students learning about gravitational waves.

## Features

Features of `Riroriro` include:

- Simulating the gravitational waveform signal from a binary merger of two black holes, two neutron stars or a black hole and a neutron star and outputting the data of this signal in terms of frequency and strain amplitude.
- Using a gravitational wave output and given a detector noise spectrum (such spectra are made publicly available by LIGO), calculating the signal-to-noise ratio (SNR) of the signal at a given distance assuming optimal alignment.
- Calculating the horizon distance (maximum distance at which an event could be observed) for a gravitational wave model and a given detector.
- Given the optimal-alignment SNR of an event, evaluating its detectability, the probability that the event would be detected with a SNR above the commonly used threshold of 8, if the alignment would be arbitrary. These results could then be combined with population synthesis calculations to estimate how many of the predicted mergers would be detected.

In addition, to help users get started with `Riroriro`, we have created Jupyter Notebook tutorials (see here).

## Research

`Riroriro` has been used for research in conjunction with BPASS, a suite of computer programs that simulates the evolution of a population of binary and single-star systems from a wide range of initial conditions and predicts their electromagnetic spectral emission (Eldridge et al., 2017; Stanway & Eldridge, 2018). There is also a Python interface for BPASS called `Hoki` (Stevance et al., 2020). This research took rates of formation of merging systems from BPASS and then evaluated the detectability of the gravitational wave signals from those systems using `Riroriro`. This was done to obtain predictions of the rates at which gravitational waves of different types would be expected to be observed, which can then be directly compared to those events found by the LIGO/Virgo gravitational wave observatories (Ghodla et al., forthcoming).

## Acknowledgments

HFS and JJE acknowledge support from the University of Auckland and also the Royal Society of New Zealand Te Apārangi under the Marsden Fund.